\documentclass{aastex62}

\accepted{\today}
\submitjournal{ApJS} 

\shortauthors{Goodarzi et al.}

\begin{document}

\title{Flare Activity and Magnetic Feature Analysis of the Flare Stars}
\correspondingauthor{Hadis Goodarzi}
\email{goodarzi@ipm.ir, hadisgoodarzy@yahoo.com}

\author[0000-0001-9128-2012]{Hadis Goodarzi}
\affil{School of Astronomy Institute for Research in Fundamental Sciences (IPM) \\
P.O. Box 19395-5746, Tehran, Iran\\}

\author{Ahmad Mehrabi}
\affiliation{Department of Physics, Bu Ali Sina University \\
65178, 016016, Hamedan, Iran \\}

\affiliation{School of Astronomy Institute for Research in Fundamental Sciences (IPM) \\
	P.O. Box 19395-5746, Tehran, Iran\\}

\author{Habib Khosroshahi}
\affiliation{School of Astronomy Institute for Research in Fundamental Sciences (IPM) \\
	P.O. Box 19395-5746, Tehran, Iran\\}


\author{Han He}
\affiliation{CAS Key Laboratory of Solar Activity\\ National Astronomical Observatories, Chinese Academy of Sciences\\
Beijing, Peopleʼs Republic of China}

\begin{abstract}

We analyze the light curve of 1740 flare stars to study the relationship between the magnetic feature characteristics and the identified flare activity. Coverage and stability of magnetic features are inspired by rotational modulation of light curve variations and flare activity of stars are obtained using our automated flare detection algorithm. The results show that (i) Flare time occupation ratio (or flare frequency) and total power of flares increase by increasing relative magnetic feature coverage and contrast in F-M type stars (ii) Magnetic feature stability is highly correlated with the coverage and the contrast of the magnetic structures as this is the case for the Sun (iii) Stability, coverage and contrast of the magnetic features, time occupation ratio and total power of flares increases for G, K and M-type stars by decreasing Rossby number due to the excess of produced magnetic field from dynamo procedure until reaching to saturation level. 

\end{abstract}


\section{Introduction} \label{sec:intro}
Solar and stellar flares are generally interpreted as manifestation of magnetic energy released in short time scales. These events occur mostly in stars with convective envelope and result from reconnection of magnetic field in stellar atmosphere \citep{1989Sph...121...299}. Reconfiguration of magnetic field geometry to lower state causes a release of energy and electromagnetic radiation in almost all wavelengths. Furthermore, the released energy accelerate particles which affect stellar atmosphere and possibly characteristics of nearby planets, including their habitability. Stellar flares are expected to occur with the same mechanism of solar flares, however a wider range of energy radiation and time duration casts doubt on the general solar analogy \citep{2010araa...48...241}. 
  
The Kepler mission has given the possibility of time-domain study on stellar magnetic activity and their flare analysis without interruption \citep{Borucki2010,Koch2010,Walkowicz2011}. The main task of this mission was to find Earth-size planets orbiting in the habitable zones of solar-like stars. This is affected by several factors including flare rate of the host star, which open a new era in stellar magnetic activity studies. The large number of stars observed by the Kepler and the high precision photometry of the data provides the opportunity for studying the properties of stars which extend our stellar knowledge dramatically. 

For instance, using 120 days of Kepler observations, \citet{Maehara2012} studied super-flares amplitude and energy on 148 solar type stars and concluded that the presence of a hot Jupiter is not correlated with super-flare on solar type stars. This study followed by \cite{Shibayama2013} which selected 279 G-type dwarfs from 500-day observation to study frequency of super-flare on solar type stars and evaluated the probability of such events on the Sun once in 2000 years. Also \citet{Notsu2013} used a sample of the same duration data (Q0-Q6) of solar type stars 
with super-flare activity and performed spot model analysis to estimate rotation period and its relationship with flare energy and frequency. \cite{Wu2015} estimated power-law index of flare frequency as a function of energy for 77 G-type stars using long-cadence data and concluded that the power-law indices can vary significantly among individual stars.

In addition to long-cadence data of Kepler instrument, short-cadence data have also been used to  analyze the properties of stellar flares with an increased time resolution and accuracy. For example, \cite{Hawley2014} studied 5 M-dwarf stars using short-cadence data to include flares with duration lower than 30 minutes and found no correlation between flare occurrence or energy with starspot phase. Also \cite{Balona} analyzed 209 flare stars in short-cadence mode and found that about one-third of flares have complex decay shapes. 
 
In order to make a catalog of flare stars, \cite{Davenport2016} utilized every available Kepler light curve (Q0-Q17) and found 4041 flare stars in the sample. He considered power law in flare occurrence and proposed saturated flare regime for fast rotating stars.
More recently, \citet{VanDoorsselaere2017} used Q15 of long cadence data and detected 6662 flare stars which extended previous flaring A-type list and also added more flare stars to the giant group. They also found flare stars from all spectral types which showed positive correlation between rotation rate and flare parameters such as the frequency and the strength. 

Kepler data have also being used to deduce and study magnetic activity of the stars from the light curve variation which is due to the dark and bright magnetic structures on the stellar surface \citep{Mehrabi2017,Giles2017,He2015}. On the Sun, bright faculae and dark sunspots are the main contributors to the solar light curve variations \citep{Koutch77,Koutchmy&stell78,Foukal&Lean86}. Sun-like stars can be divided into two groups, facula-dominated and spot-dominated, in which their brightness fluctuations are correlated or anti-correlated in time with magnetic activity of the stars, meaning that the long-term variations of light curve are driven by bright faculae or dark sunspots, respectively \citep{Montet17}. ُThis stellar activity studies are important not only in stellar community but also in the exoplanet field to decrease false detections and possible errors in determining planet characteristics \citep{Giles2017}.

\cite{He2018} selected three solar type stars and considered the relationship between long-term variation of magnetic features and flare activity. For this purpose, they defined two magnetic proxies and three flare indexes to indicate different characteristic of magnetic features and flares quantitatively.  \cite{Mehrabi2019} followed the work by \cite{Mehrabi2017} on a sample of M-type stars and analyzed their magnetic activity properties which found higher value of light-curve fluctuations for stars with higher rotation rate. 

In this paper we select the largest sample of flare stars, that is the combination of flare stars from catalogs available in \citet{Davenport2016} and \citet{VanDoorsselaere2017} with determined rotation period from \citet{McQuillan2014}. Magnetic and flare activity of the sample are studied from the light curve variations of 1740 flare stars from spectral type F, G, K and M.
We develop an automated detection algorithm to find flares and calculate magnetic proxies and flare indexes for flaring stars during approximately 4 years of observations. Then, we perform statistical study on flare activity and its relationship with different parameters such as magnetic feature stability, coverage and contrast, Rossby number and spectral type of the stars.

We present data selection and processing method in section \ref{sec:process} and describe our automated routine for detection of flares in section \ref{sec:flrdet}. In section \ref{sec:param}, magnetic proxies and flare indexes are introduced and the result of magnetic feature analysis is presented in section \ref{subsec:Magnetic}. We explain the properties of flare activity in section \ref{subsec:flare} and then, in section \ref{subsubsec:corr} the result of correlation analysis
are described. We discuss the results and give our conclusions in section \ref{sec:conclusion}.  

\section{Data Selection and processing} \label{sec:process}

Kepler data have two modes of data recording, long cadence and short cadence data with 29.4 minutes and 1 minute time intervals, respectively \citep{Jenkins2010a}. Most of the Kepler data have been recorded in long cadence mode which is appropriate data sample for statistical analysis on flare events, specially the ones with higher energy amplitudes (or super-flares).

We use the PDC (Pre-search Data Conditioning) flux data that are the result of PDC module of the Kepler science processing pipeline \citep{Jenkins2010b, Jenkins2017} in which the first steps of data corrections have been done on the level zero data \citep{Smith2012, Stumpe2012, Stumpe2014}. Long cadence data from quarters 2-16 (data release 25) are selected, data from Q0-Q1 and Q17 are neglected because of their shorter time duration. The physical parameters of the Kepler stars such as effective temperatures and surface gravities have been obtained using multi-color photometry, and was listed in the initial Kepler Input Catalog \citep{Brown2011}. However, this stellar parameter values were revised based on broadband photometry, asteroseismology and other techniques which are more accurate than the previous catalog \citep{Huber14,Huber et al.14}.
We select our targets in which their magnetic activities are confirmed from combination of two flare activity catalog in \cite{Davenport2016} and \cite{VanDoorsselaere2017}. From these two flare star catalogs, we select targets in which their rotation period have been determined by automated auto-correlation method in the catalog of \cite{McQuillan2014}. So, our sample includes stars with temperature below 6500, since this threshold value selected by \cite{McQuillan2014} as upper boundary to ensure that all of the stars have convective envelope and spin down during their lifetime. \cite{Berger18} derived precise radii and evolutionary stated of the Kepler stars using Gaia Data
Release 2 and we use table1. of this paper to remove the binary candidates and subgiants. 
Final sample contains 1740 single, main-sequence flare stars from different stellar type (F, G, K and M-type) and rotation periods ranging from several hours to 45 days. M-type stars have lower population in comparison with the other three stellar types. 

\begin{figure}
	\epsscale{0.88}
	\plotone{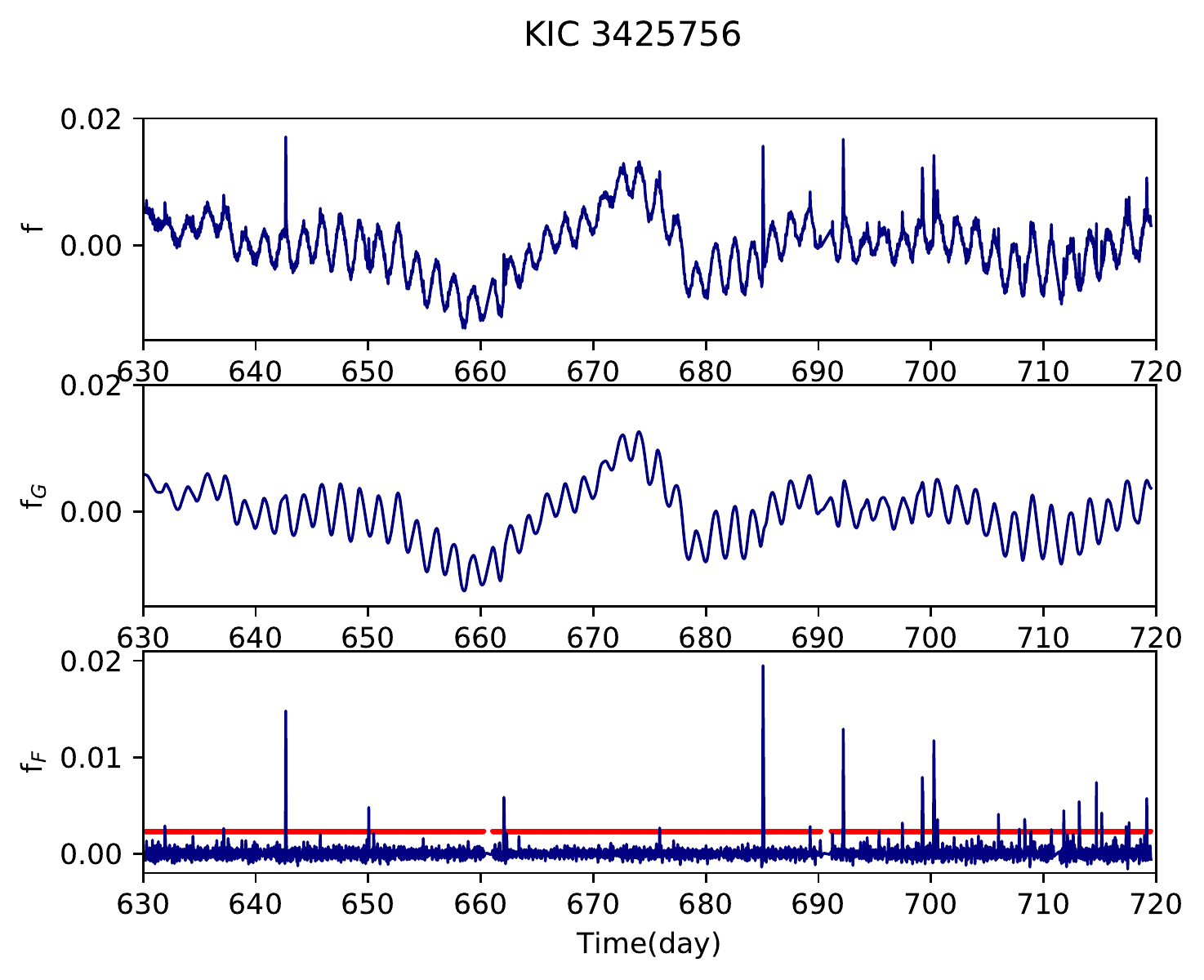}
	\caption{Normalized flux of the KIC 3425756 (top), normalized gradual component (middle) and normalized flare component (bottom). The height of red horizontal line in the bottom panel shows threshold value for detecting super-flares and its length shows valid observation time, the gaps are due to spacecraft data down-link.} 
	\label{Fig1}
\end{figure}

\section{Flare detection algorithm} \label{sec:flrdet}
In order to extract flare component, we developed an automated routine to distinguish flares. The step by step procedure of the flare detection algorithm is described as follows.

1. The initial step is to compute the gradual component of the flux variation ($F_{g}$) which is due to the rotation modulation of magnetic features. We first filter patched data points with low-pass filter and the cutoff frequency depending on rotation period of the star which has been selected empirically after numerous visual inspection in each rotation period span. This led to elimination of flare spikes.

2. The missing data points left by removal of prominent flare in the previous step were patched with linear interpolation of full data points. By repeating this procedure, gradual component were deduced ($F_{g}$). 

3. The original flux (F) and the gradual component ($F_{g}$) are normalized to the median value of gradual component, which have been referred to, after normalization, as “f” and “$f_{g}$ ”, respectively (Figure \ref{Fig1}, top and middle panel).

4. Final flare component is the difference of the normalized flux and the normalized gradual component $f_{F}$= f -$f_{g}$ (Figure \ref{Fig1}, bottom panel).

5. To infer the flare candidate automatically, three times the standard deviation of running difference ($3*\sigma$) are chosen as the threshold value of flares (the red horizontal line in bottom part of Fig.\ref{Fig1}) to avoid spurious sources and to identify prominent flares or super flares \citep{Wu2015}. We checked visually our automatic flare detection method to insure that it works well for different light curves with various flare magnitude and noise levels. Using the mentioned procedure, 590452 flares are detected on 1740 flare stars.

\section{Magnetic feature and flare parameters} \label{sec:param}
We follow the definition of magnetic proxies in the paper by \cite{He2015} and \cite{He2018} to describe quantitatively the magnetic activity characteristic, including stability and relative coverage of the magnetic features. However, some modifications have been performed in order to take into account the contrast between stellar photosphere, magnetic features and flaring region which depends on their temperature. 
  
The first parameter is the auto-correlation index ($i_{AC}$) that describes the degree of periodicity of the light curve and estimates the stability of the magnetic features. In this way, long lived magnetic features on the stellar disk cause stable variation shape of the light curve that leads to higher value of auto-correlation index \citep{He2015,Giles2017}. 
For an observed time series with N data points $\{X_t, t=0, 1, \ldots, N-1\}$ , the auto-correlation index is defined as below,
\begin{equation}\label{equ:acf}
\rho(h)=\frac{\sum_{t=0}^{N-1-h} (X_{t+h}-\overline{X})(X_t-\overline{X})}{\sum_{t=0}^{N-1} (X_t-\overline{X})^2},
\qquad 0\leqslant h \leqslant N-1,
\end{equation}
\begin{equation}\label{equ:iac}
i_{\rm AC}=\frac{2}{N} \int_{0}^{N/2} |\rho(h)| dh,
\end{equation}
Where $\overline{X}$ is the mean value of $X_t$ and $\rho(h)$ is auto-correlation coefficient at the time lag \textit{h}. The value of \textit{N/2} in the integration limits is the maximum possible value which allows the entire light curve to be included in the calculations. As the length of each quarter is 90 days, we selected stars with rotation period below 45 days to cross correlate each light curve with itself because higher rotation periods do not give factual value of $i_{AC}$. Maximum possible value for auto-correlation index is approximately 0.48 that belongs to regular sinusoidal wave \citep{He2015}.  
 
The second parameter is the effective range of light curve variation ($R_{eff}$) which provides an estimate for the relative magnetic features coverage and contrast on the stellar disk. The major part of the light curve variations arise from stellar spots \citep{Namekata19} and the contrast between the photosphere and the spots depends on the effective temperature of the stars ; in stars with higher temperature, spots tend to show a higher contrast with respect to the surrounding photosphere  \citep{Berdyugina2005}. Thus we estimated spot temperature ($T_{spot}$) using Fig.7 in \cite{Berdyugina2005} and its extrapolation near to 6700K for stars with higher temperature in our sample.
For $X_{t}$ time series $R_{eff}$  is calculated as below, 
\begin{equation}\label{equ:x_t}.
x_t=\frac{X_t-\widetilde{X}}{\widetilde{X}},
\qquad t=0, 1, \ldots, N-1,
\end{equation}
\begin{equation} \label{equ:reff}
R_{\rm eff}=2\sqrt{2} \cdot x_{\rm rms} \cdot [{1-(\frac{T_{spot}}{T_{phot}})^4}] =2\sqrt{2} \cdot \sqrt{\frac{1}{N} \sum_{t=0}^{N-1} x_t^2} \cdot [{1-(\frac{T_{spot}}{T_{phot}})^4}],
\end{equation}
where $\widetilde{X}$ is the median value of $X_t$, $x_{rms}$ is root mean square of the time series and the $2\sqrt{2}$ coefficient gives the effective range. We add the term of $[{1-(\frac{T_{spot}}{T_{phot}})^4}]$ to take into account the contrast between photosphere and the spots \citep{Notsu2013}, where $T_{phot}$ is temperature of the photosphere and we use $T_{eff}$ instead from the Kepler input catalog.  
Stronger magnetic field in the stellar spots can inhibit more effective transport of the heat from interior parts, which results in a darker surface in comparison with the surrounding photosphere and higher contrast. Therefore, $R_{eff}$ reflects not only the relative size of the magnetic features on the stellar surface, but also gives an estimation of their magnetic field strength. Nevertheless, it should be noted that the amplitude of rotational modulation gives an estimation of the largest magnetic features on the stellar surface. So, in the case of several small magnetic structures at different longitudes, $R_{eff}$ would not be a good indicator of the total magnetic flux.

In addition to the magnetic proxies, three parameters have been used to describe the flare characteristics. The first parameter is time occupation ratio of flares, that is the total time occupied by flare spikes ($ T_{flare} $) divided by the valid observation time ($ T_{obs} $),

\begin{equation} \label{equ:Rflare}
R_{\rm flare}=T_{\rm flare}/T_{\rm obs}
\end{equation}

$R_{\rm flare}$ is a measure of the flare occurrence frequency and takes higher value when the star produce flare more often.

The second flare index is the total power ($P_{\rm flare}$) which is the ratio of total energy ($U_{\rm flare}$) released by all diagnosed flare spikes ($f_{s}$) to the valid observation time,

\begin{equation} \label{equ:Pflare}
P_{\rm flare}=U_{\rm flare}/T_{\rm obs}.
\end{equation}

In order to calculate $U_{\rm flare}$, we assume the star as a black-body radiator that emits energy in its effective temperature, therefore, the bolometric flare luminosity ($ L_{flare} $) is calculated from temperature and area of the flares ( $ T_{flare} $ and $ A_{flare} $, respectively) using the following equation \citep{Shibayama2013},

\begin{equation} \label{equ:Lflare}
L_{\rm flare} = \sigma_{\rm SB} T_{\rm flare}^4 A_{\rm flare}~,
\end{equation}

where $\sigma_{SB}$ is the Stefan-Boltzmann constant and we assume $ T_{flare} $ to be 9000K as the spectral energy distribution (SED) of flares can be fitted by the black-body spectrum with this temperature \citep{HawleyFisher92, Kretzschmar2011}. Flare area can be estimated using following equation, 

\begin{equation}\label{equ:area}
A_{\rm flare} = f_{\rm S} \pi R^2 \frac{\int R_{\lambda}B_{\lambda(T_{\rm eff})}d\lambda}{\int R_{\lambda}B_{\lambda(T_{\rm flare})}d\lambda}~.
\end{equation}

$f_{s}$ is the flare spike amplitude of the light curve after detrending, $B_{\lambda}(T)$ is Planck function, $R_{\lambda}$ is the response function of the Kepler instrument\footnote[1]{https://keplergo.arc.nasa.gov/CalibrationResponse.shtml.} and \textit{R} is the stellar radius. Then, total bolometric energy of flare is an integral of $L_{flare}$ during the flare event \citep{Shibayama2013},

\begin{equation}\label{equ:Uflare}
U_{\rm flare} = \int_{\rm flare} L_{\rm flare}(t)~dt~.
\end{equation}

The last parameter is the averaged flux magnitude of flares ($M_{\rm flare}$) that is defined by the equation,

\begin{equation} \label{equ:Mflare}
M_{\rm flare}=\frac{1}{T_{\rm flare}} \int f_{\rm S}(t) dt=U_{\rm flare}/T_{\rm flare}.
\end{equation}


Since we estimate the bolometric energy of the flares using their absolute area (equation \ref{equ:area}), consequently the value of the total power and magnitude of flares are absolute. Also from the above equations we see that flare indexes are related to each-other by the following equation,
 
\begin{equation} \label{equ:RPMflare}
P_{\rm flare}=M_{\rm flare} . R_{\rm flare}.
\end{equation}

\section{Results}
Magnetic proxies such as auto-correlation index ($i_{AC}$), effective range of light curve fluctuations ($R_{eff}$) and also flare indices including time occupation ratio ($R_{\rm flare}$), total power of flares ($P_{\rm flare}$) and averaged flux magnitude of flares ($M_{\rm flare}$) were computed for quarters Q2-Q16 of 1740 stars separately and the mean of each parameter over 15 quarters were calculated, see table \ref{tab:parameters}. In this section we consider the general behavior of these parameters as a function of Rossby number (Ro) which is the ratio of rotation period to the convective turn-over time. It has been shown that when magnetic activity of the star is being considered, Rossby number is a better indicator than rotation period alone \citep{Montesinos01}. We also study variation of magnetic proxies and flare indexes toward each-other to have general inspection about flare and magnetic activities of selected flare stars.

Based on effective temperature of the stars we divide our sample into four spectral types; F-type stars (6000$<T_{eff}<$6700) indicated with blue dots, G-type stars (5200$<T_{eff}<$6000) indicated with green dots, K-type stars (3700$<T_{eff}<$5200) indicated with yellow dots and M-type stars ($T_{eff}<$3700) indicated with red dots, see for example, Fig \ref{Fig4}. 

\floattable
\begin{deluxetable}{lccccccccccccccc}
	\tablecaption{Prarameters of Kepler flare stars. This table is available in its entirety (1740 stars) and machine-readable form in the online journal.
		 \label{tab:parameters}}
	\tablehead{
		\colhead{KIC} 
		&&  \colhead{$T_{\rm eff}$\tablenotemark{1}}  & \colhead{$P_{\rm rot}$\tablenotemark{2}} & 
		  \colhead{$i_{\rm AC}$}  & \colhead{$R_{\rm eff}$} & \colhead{$R_{\rm flare}$}
		&  \colhead{$P_{\rm flare}$}  & \colhead{$M_{\rm flare}$} & \colhead{Thr\tablenotemark{3}} & \colhead{$N_{\rm f}$\tablenotemark{4}} \\
		&&                   (k) &                        (day)&              & $(10^{-2})$&
		$(10^{-2})$& $(10^{22})$        &$(10^{24})$  &$(10^{-2})$                         
	}
	\startdata
	757099  &&  5519 & 0.367 & 0.344 & 3.580 & 0.508 & 7.664  & 15.176 & 0.307  &  204 && \\
	892376  && 3973 & 1.532  & 0.269 & 0.4861& 0.966 & 0.397  & 0.453  & 0.175  &  319 && \\ 
	1025986 && 5841 & 9.724  & 0.250 & 1.376 & 0.514 & 0.079  & 0.147  & 0.013  &  207 && \\
	1028018 && 5930 & 0.621  & 0.297 & 4.691 & 0.739 & 2.575  & 3.474  & 0.199  &  272 && \\	
	1434277 && 5842 & 10.446 & 0.300 & 1.182 & 0.833 & 0.178  & 0.225  & 0.018  &  333 && \\
	1569863 && 3591 & 13.321 & 0.361 & 0.8828& 1.116 & 0.149  & 0.143  & 0.288  &  433 && \\
	1570924 && 4961 & 3.234  & 0.327 & 9.05  & 1.908 & 1.547  & 0.958  & 0.229  &  490 && \\
	1572802 && 3878 & 0.374  & 0.334 & 3.029 & 1.452 & 1.843  & 1.494  & 0.874  &  470 && \\
	1721614 && 6221 & 1.453  & 0.255 & 2.554 & 1.012 & 0.924  & 0.918  & 0.082  &  359 && \\
	1722506 && 4334 & 10.653 & 0.409 & 2.106 & 0.991 & 0.584  & 0.630  & 0.273  &  210 && \\
	\enddata
	\tablenotetext{1}{From the Kepler Input Catalog \citep{Huber et al.14,Huber14}}
	\tablenotetext{2}{From the rotation period catalog derived by \cite{McQuillan2014}.}
    \tablenotetext{3}{Threshold value of flare detection on the normalized flare component, see section \ref{sec:flrdet}.}
    \tablenotetext{4}{Number of flares.}
\end{deluxetable}

\subsection{Magnetic features stability, coverage and contrast} \label{subsec:Magnetic}
\begin{figure}
	\epsscale{01.2}
	\plotone{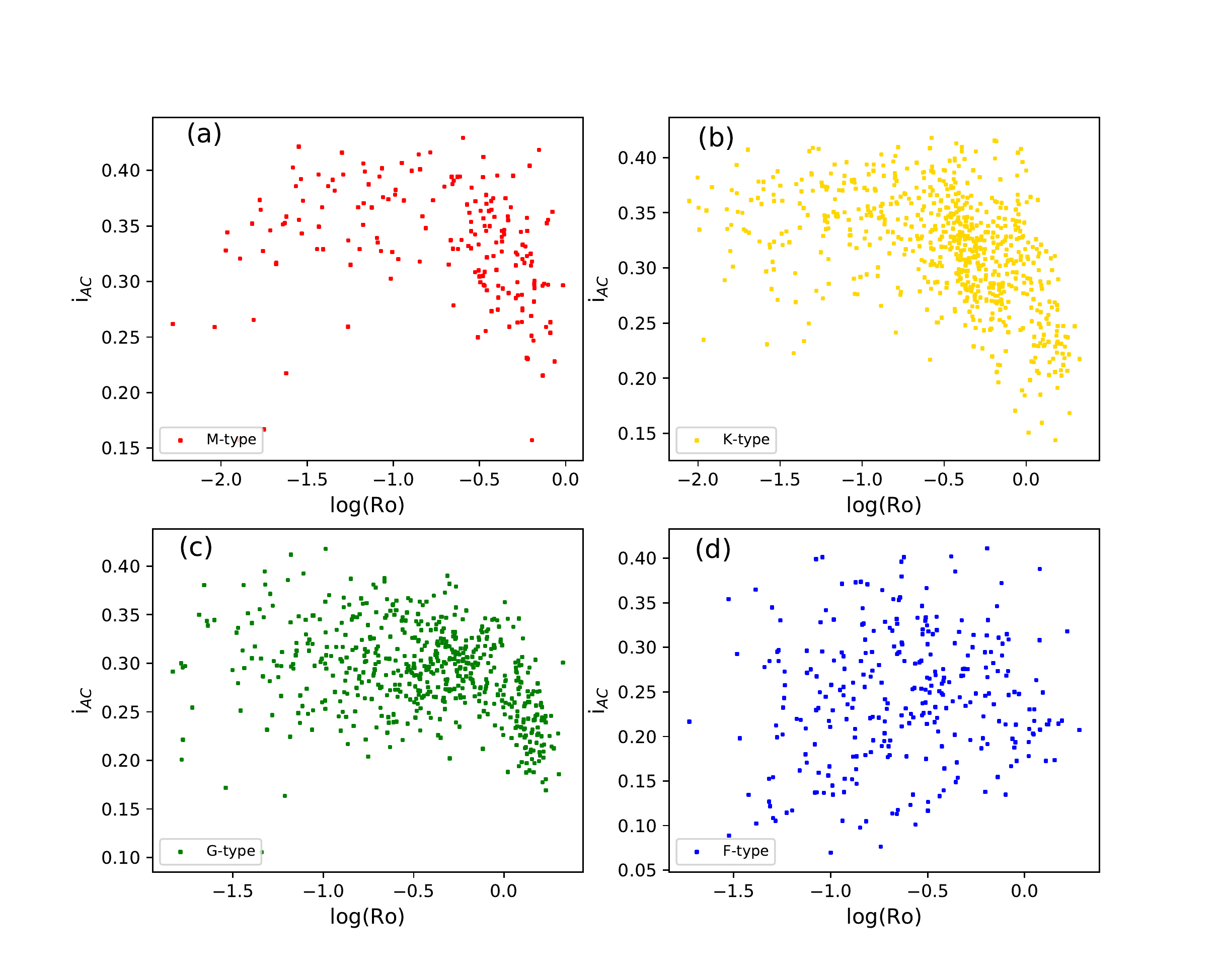}
	\caption{Auto-correlation index ($i_{AC}$) as a function of Rossby number for different spectral classifications: M-type stars (top left), K-type stars (top right), G-type stars (bottom left) and F-type stars (bottom right). }
	\label{Fig2}
\end{figure}

Figure \ref{Fig2} shows the auto-correlation index $i_{AC}$ for different spectral types as a function of Rossby number. Auto-correlation index increases with decreasing Rossby number in stars from G, K and M spectral types until reaching to saturation level. However in F-type stars with higher effective temperature, $i_{AC}$ does not change with Ro and scatter everywhere.
Increasing magnetic feature stability with decreasing Ro in G-M type stars can be attributed to the more effective magnetic field production from dynamo procedures at lower Rossby numbers.
Lowest values of magnetic feature stability ($i_{AC}<0.1$) belong to F-type stars. This can arise from the fact that the spots may survive shorter during time on hotter stars due to the higher vigor of convection and turbulent diffusivity which leads to increasing the rate of spot decay \citep{Simon&Leighton1964,Giles2017} .   

\begin{figure}
	\epsscale{1.3}
 \plotone{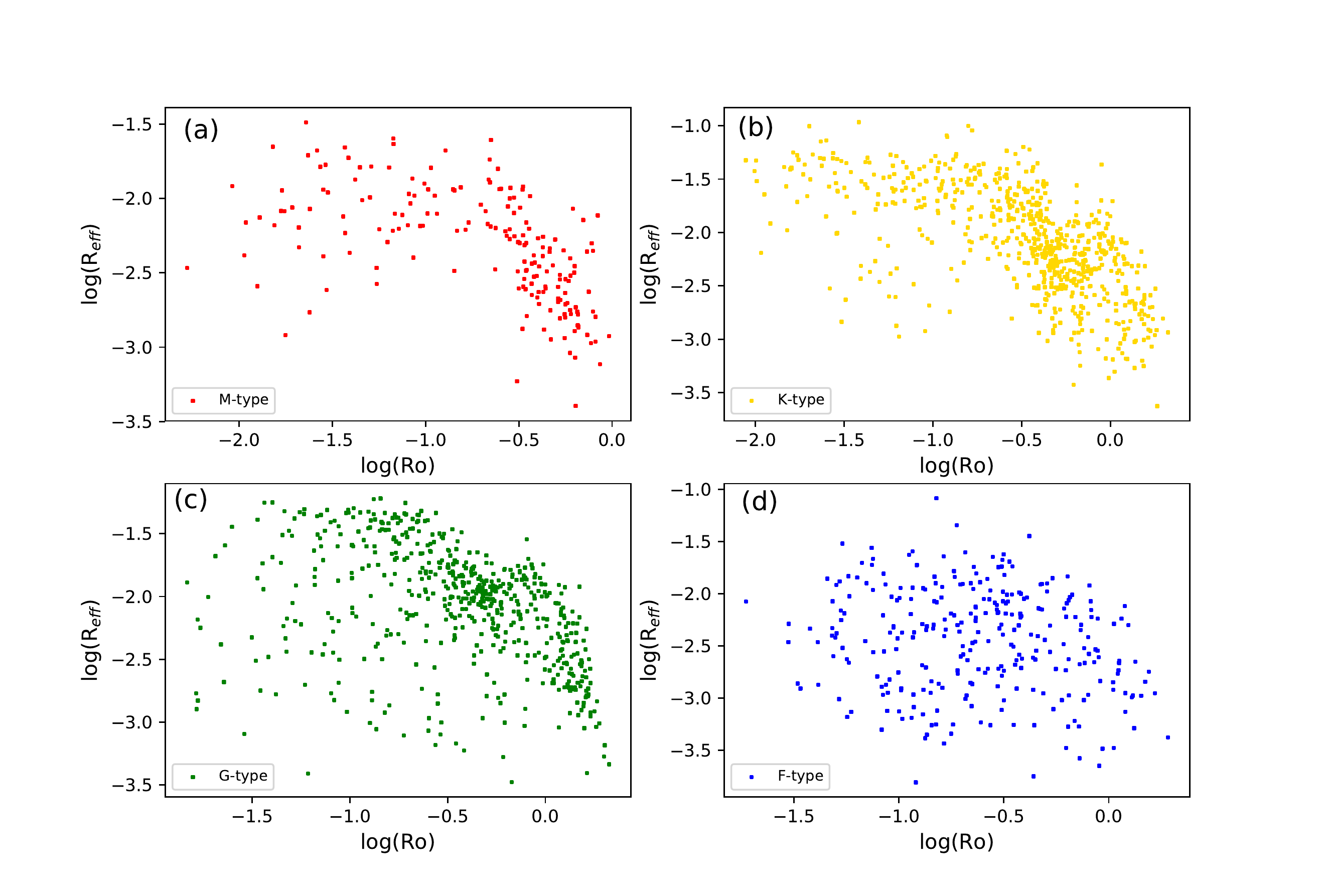}
	\caption{Effective range of light curve fluctuation ($R_{eff}$) as a function of Rossby number for different spectral classifications: M-type stars (top left), K-type stars (top right), G-type stars (bottom left) and F-type stars (bottom right). }
	\label{Fig3}
\end{figure}

The next magnetic proxy, $R_{eff}$ measures the effective range of the light curve fluctuations and shows the relative coverage of magnetic features together with the contrast produced by them. In Figure \ref{Fig3}, $R_{eff}$ is shown as a function of Rossby number for the four spectral types. In M, K and G-type stars relative coverage and contrast of magnetic features increases by decreasing Ro, until reaching to saturation level around $log Ro\sim$-0.5 and does not increase more after that. The anti-correlation between the rotation period and spot coverage has been confirmed before, specially in cooler stars \citep{Candelaresi14}. For example, \cite{Notsu2019} concluded that for Sun-like stars, dependence of starspot area on rotation period is weaker when the star is young and rapidly rotating. However, when the star becomes older and its rotation slows down, the maximum area of the starspot have a steep decreasing after a certain point.  
In hotter F-type stars there is not a clear trend between the rotation rate and the amplitude of rotational variations (panel d of Figure \ref{Fig3}) and this was verified also by \cite{McQuillan2014}

Saturation of magnetic field has been reported in different types of activity indicators \citep{Vilhu84} and means that at high rotation rates (or small Ro), all activity gauges saturate and do not increase above a certain value, regardless of higher rotation rate \citep{James2000}. 

One way to measure the magnetic flux in order to study the saturation phenomena is to focus on the magnetically sensitive lines. By using this technique, \cite{Reiners2009} studied M-type stars and concluded that both strength and filling fraction of magnetic field saturate at $log Ro\sim$ -1.  
However, in rapid rotators earlier than M-types, the high surface rotation velocities required for activity saturation prevents the measuring of magnetic fluxes and make it impossible to see saturation phenomena using magnetically sensitive lines \citep{Reiners2009}.
Here, by analyzing the light curve fluctuations, we find that the saturation regime occurs on M, G, K type stars in magnetic feature coverage and contrast, which can be manifestation of filling fraction and strength of magnetic field.

As we mentioned before, the amplitude of rotational brightness modulations is higher for one big magnetic feature in comparison with multiple small structures at different longitudes, and $R_{eff}$ may not be a good indicator of the total magnetic flux. So, the large scatter of magnetic field production under the maximum value specially at low Rossby numbers can be due to this fact and has been also reported by \cite{Reiners2009}.

\begin{figure}
	\epsscale{0.8}
	\plotone{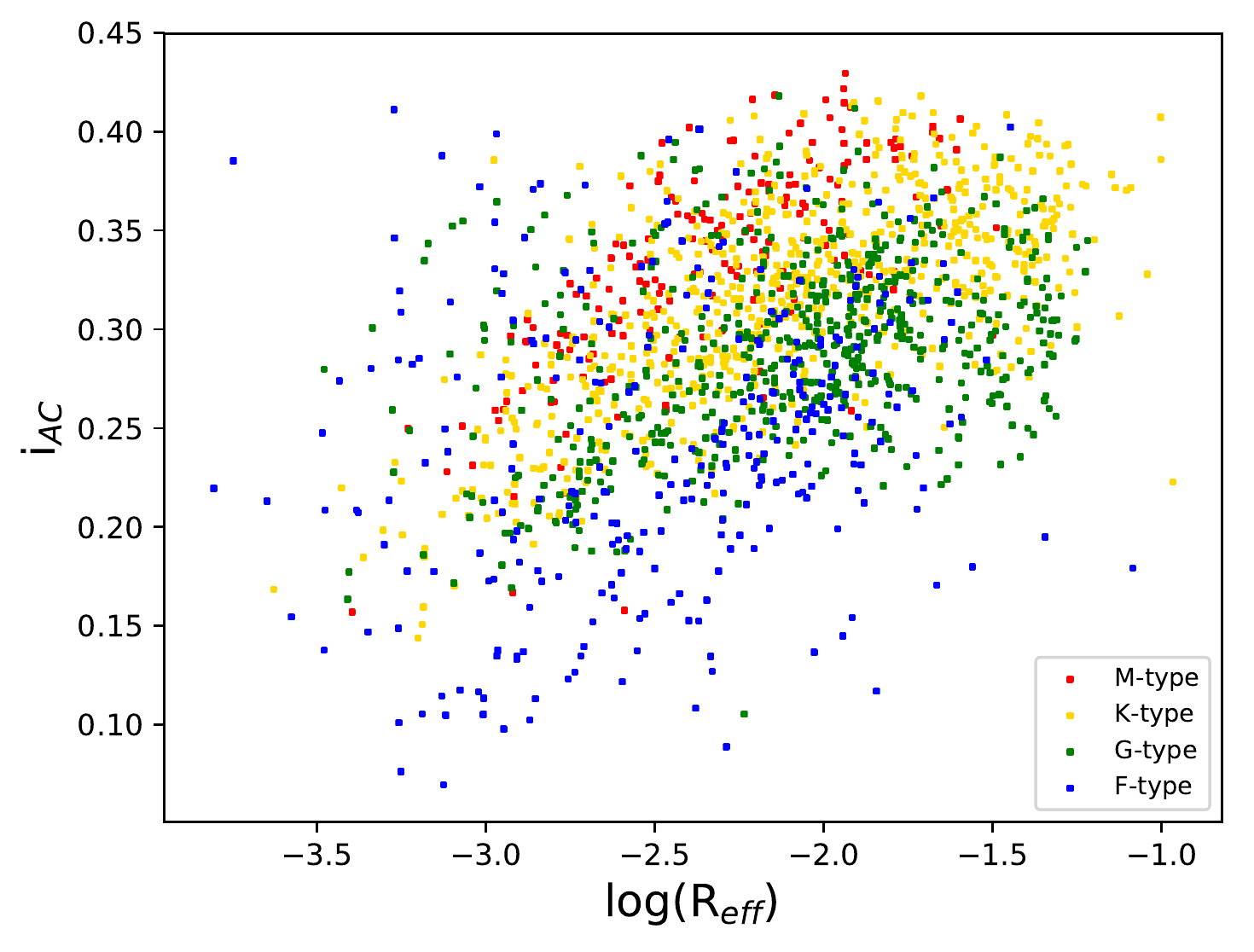}
	\caption{Auto correlation index ($i_{AC}$) versus effective range of light curve variations ($R_{eff}$), colors indicate spectral type of the stars.}
	\label{Fig4}
\end{figure}

In Figure \ref{Fig4}, we see the variation of $i_{AC}$ as a function of $R_{eff}$ for different spectral types. Colors indicate effective temperature of the stars or their spectral classifications. Although F-type stars have the higher dispersion than the others, the stability of magnetic features on all of the four spectral types increases with increasing $R_{eff}$. In other words, magnetic features with larger relative coverage and the more powerful ones which produce higher values of contrast on the stellar surface live longer than the smaller and lower magnetic field structures.

Before, size-lifetime relation has been confirmed from data analysis of stars that have spots on their surfaces \citep{Hall94,Bradshaw2014}. For example, \cite{Giles2017} used two samples of stars with rotation period around 10 and 20 days and found the same relationship between lifetime and spotsize. However, they concluded that hotter stars with $T_{eff}$ greater than 6200K have smaller range of spot size and lifetime than the cooler stars (see their Fig.5) which is inconsistent with our Figure \ref{Fig2} and \ref{Fig3}, in which F-type stars can have values of $i_{AC}$ and $R_{eff}$ as high as the other three spectral types. This can arise from different sample selection, since they did not include stars with rotation period below 10 days. At higher Rossby numbers, our results of $R_{eff}$ are consistence with each other, see panel (d) of Figure \ref{Fig3} that at $log Ro>$ 0, average value of $R_{eff}$ is lower than G and K type stars in panel (b) and (c) of this figure.

\cite{Namekata19} estimated the area of stellar spot using brightness depth of local minimum from the nearby maximum on a sample of solar-type stars. They found a positive correlation between area and lifetime of the spots as well. Also they concluded that rapidly rotating stars have faster emergence and decay rate of sunspots compared to slowly rotating stars on a much smaller sample than us (lower than 50), see their Figure 5.
 
Positive relation between spot size and coverage is the case for the Sun, bigger magnetic features usually live longer than the smaller ones \citep{Gnevyshev38, Petrovay97}. We will confirm this for each individual stars by correlation analysis, see section \ref{subsubsec:corr}.

\begin{figure}
	\epsscale{1.2}
	\plotone{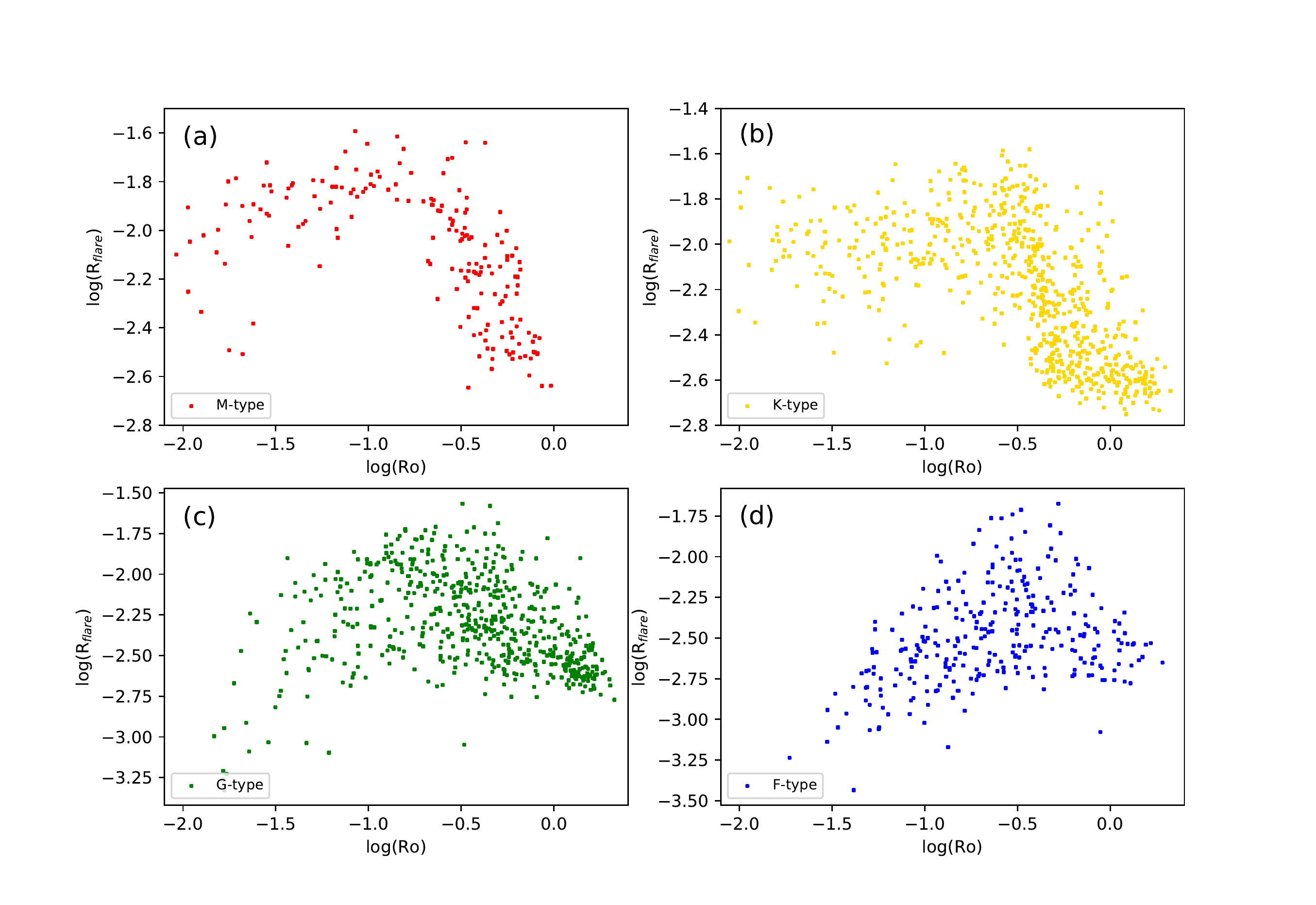}
	\caption{Time occupation ratio of flares ($R_{\rm flare}$) versus logarithm of Rossby number for different spectral classifications: M-type stars (top left), K-type stars (top right), G-type stars (bottom left) and F-type stars (bottom right).}
	\label{Fig5}
\end{figure}

\subsection{Flare characteristics} \label{subsec:flare}

\subsubsection{Time Occupation Ratio of Flares} \label{subsubsec:Rf}

Figure \ref{Fig5} shows the variation of time occupation ratio of flares or flare frequency ($R_{\rm flare}$) versus Rossby number for 1740 stars in four spectral types. Slow rotator stars with $log~Ro>-0.2$ rarely have high value of flare frequency, and this is a straightforward result from dynamo theory in which the rotation rate of the star play an important role on the magnetic field production needed for flare activity.
 
Starting from the highest value of Rossby number (close to log Ro= 0.2) in Figure \ref{Fig5}, flare frequency of the four spectral type stars increases by decreasing Ro until reaching to a maximum value around $log~Ro\sim-0.5$. At lower values of Rossby number, $R_{\rm flare}$ decreases in F and G-type stars and scatter under saturation level in cooler K-M type stars.
 
The saturation behavior of flare activity in cooler stars is in consistent with the result of \cite{Yang2019} which found approximately the same critical value of Rossby number as the boarder point separating decay and saturated regime in K and M type stars. However, \cite{Davenport2016} estimated lower threshold value (log Ro $\sim$-1.5) for stars with spectral types later than G8. This is also much smaller than the typical value of threshold Rossby number which was found using X-ray activity \citep{wright2011}. 

Different behavior of flare activities with saturation or non-dependency to rotation rate at low Rossby numbers and decrease of activity at higher Ro can be the result of different dynamo procedure at work proposed by \cite{Barnes2003}. This theory divides the main sequence stars into two groups. First group (C-class) are young and fast rotator stars that their magnetic activity depends mostly to the temperature. The second group (I-class) are slow rotators with the same dynamo as the Sun and their dynamo depends mainly to the Rossby number. In this paradigm, the stars with saturated flare activity at low Rossby numbers are from C-class and their activity depend mainly to the temperature rather than Rossby number. The stars with higher Rossby numbers are from I-class which their flare activity decreases by increasing Ro. C-class stars evolve to I-class as they age and their dynamo switch to the solar-like dynamo \citep{Barnes2003,Yang2019}.    

From the four spectral types, F-type and G-type stars have the larger variations of time occupation ratio of flares. Specially lowest value of flare frequency belong to these two spectral types. 
Nevertheless, as the background luminosity of the hotter stars are higher than cooler stars, their flare and magnetic activity may be underestimated due to the observational limitations \citep{Balona} and this should be taken into account when comparing variations of $R_{\rm flare}$ for different spectral classes. 

Figure \ref{Fig6}(a) depicts $R_{\rm flare}$ as a function of auto-correlation index $i_{AC}$, it is hard to say that $R_{\rm flare}$ increases with magnetic feature stability but a faint trend could be inferred. This can arise from indirect influence of higher relative magnetic feature coverage attributed to higher magnetic feature stabilities (see Figure \ref{Fig4}). On the other hand, long-lived magnetic features on the stellar surface does not necessarily cause to magnetic re-connection and flare production since there is a wide dispersion at high values of $i_{AC}$.

Figure \ref{Fig6}(b) shows the time occupation ratio of flare as a function of effective range of light curve variations. We see that $log~R_{\rm flare}$ is proportional with $log~R_{eff}$ and it remains valid for all of the four stellar classifications that we considered. Stars with higher value of $R_{eff}$ have larger relative magnetic feature coverage and contrast, 
which can result in higher possibility of magnetic field reconnections and the subsequent flare events.
 

\subsubsection{Total power of flares} \label{subsubsec:Pf}

Total power of flares ($P_{\rm flare}$) is the ratio of total energy released by all the identified flares to the observation time. Figure \ref{Fig7} shows $P_{\rm flare}$ as a function of Rossby number for the four group of spectral types. Fast rotating stars have wider range for power of flares and slow rotating stars with $log~Ro>-0.2$ rarely have high values of $P_{\rm flare}$. Total power of flares increases by decreasing Rossby number in M and K-type stars, until $log~Ro\sim-0.75$ which is the saturation threshold and $P_{\rm flare}$ does not increase more afterwards. This improvement and saturation behavior becomes weaker in G-type stars and completely disappear in F-type stars, see panel (c) and (d) of Figure \ref{Fig7}, respectively.

There are two explanations for saturation of magnetic activity at high rotation rates. (i) The dynamo mechanism saturates and stars can not produce magnetic field higher than a certain value. (ii) The magnetic fields proceed to be stronger at high rotation rates, but most of the stellar surface are covered by the spots and the ratio af magnetic structures area to the stellar area approaches to one, so no more emitting plasma could exists on the stellar surface. In our analysis, magnetic activity indicators such as frequency and power of flares, magnetic feature stability, coverage and contrast saturate at Rossby number around 0.18-0.30. Since all of the magnetic activity indicators depend not only on the magnetic field strength but also on their relative coverage (see section \ref{sec:param}), we can conclude that both of the strength and filling fraction of the magnetic structures have a saturation level at high rotation rates \citep{Reiners2009}.    

$P_{\rm flare}$ shows faint increase with increasing magnetic feature stability $i_{AC}$ (Figure \ref{Fig8}, panel a) specially in K and M-type stars, more stable magnetic feature could result in higher value of magnetic energy stored within them. Nevertheless, the effective range of light curve fluctuations seems to play more important role than the stability, since $P_{\rm flare}$ increases clearly with increasing $R_{eff}$ (Figure \ref{Fig8} panel b) for all spectral classifications included in this study. This can be explained by the origin of flares as the result of sudden energy release during magnetic reconnection. When relative coverage and contrast (or strenght) of the magnetic structures increase, the magnetic energy accumulated in these features increases and leads to higher values of $P_{\rm flare}$ \citep{Shibayama2013,Notsu2013}. 

\begin{figure}
	\gridline{\fig{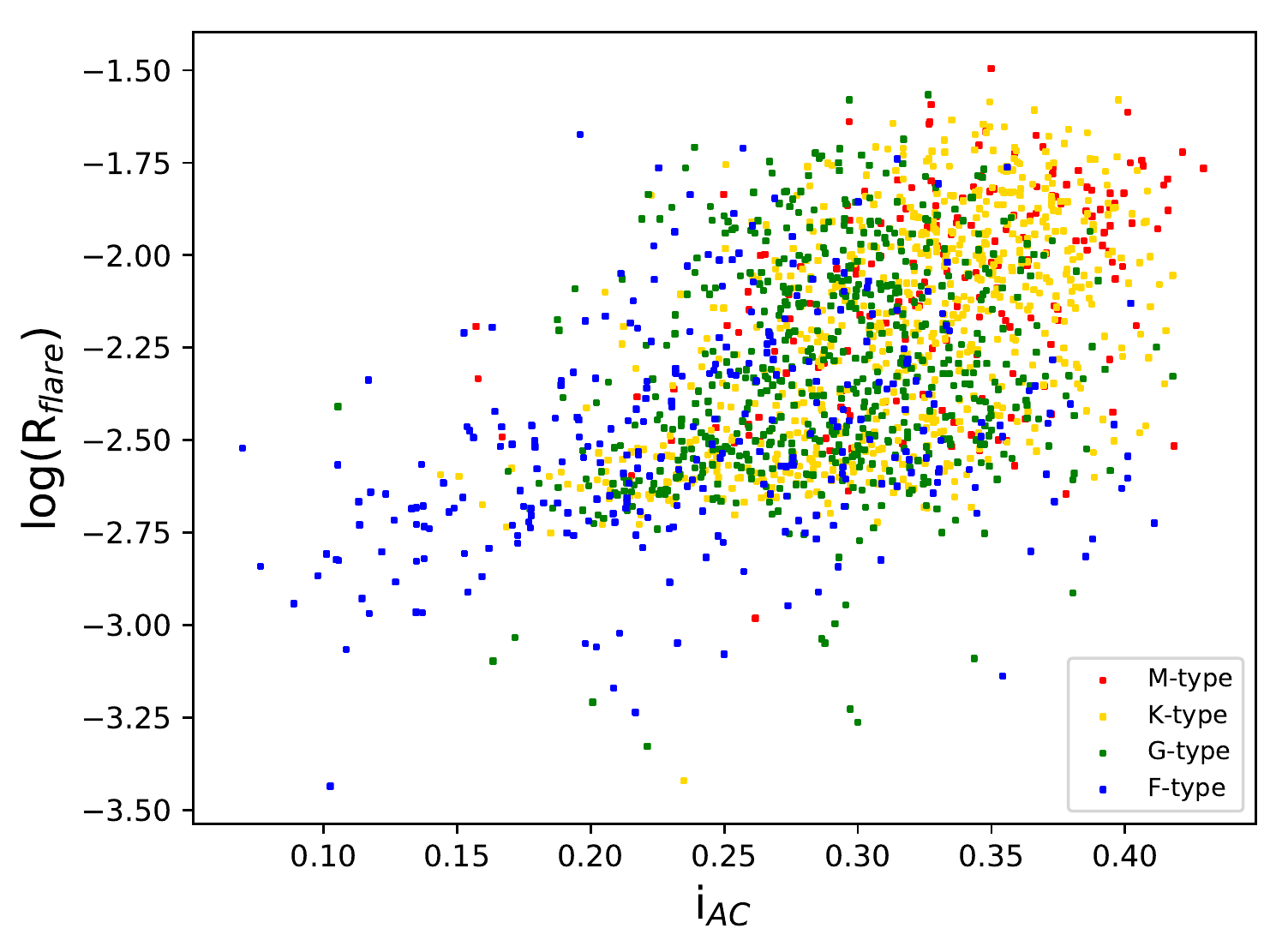}{0.51\textwidth}{(a)}
	\fig{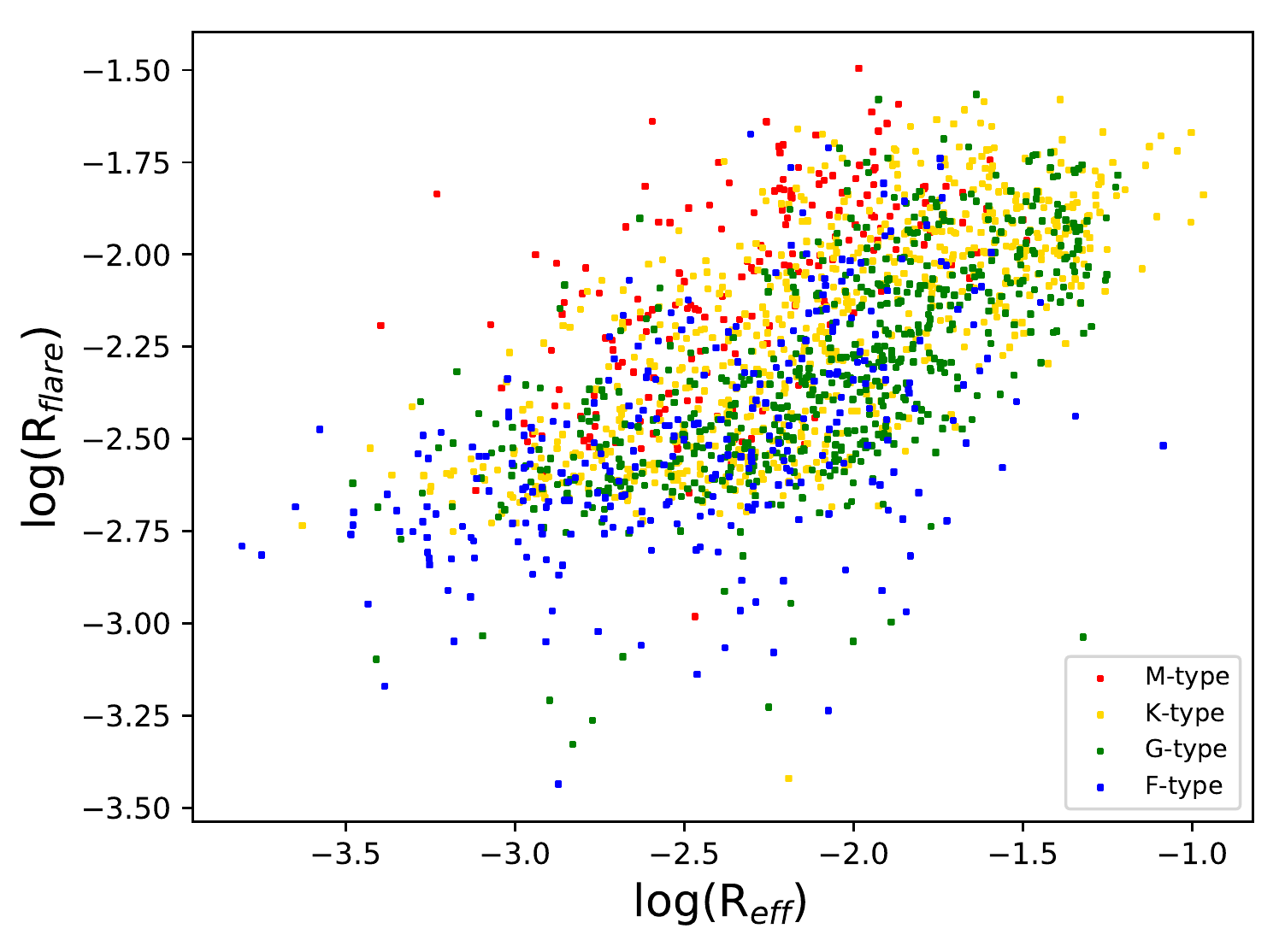}{0.51\textwidth}{(b)}}
	\caption{Logarithm of time occupation ratio of flare ($R_{\rm flare}$) versus autocorrelation index ($i_{AC}$, left) and versus effective range of light curve fluctuations ($R_{eff}$, right) for different spectral types that are indicated by colors.}
	\label{Fig6}
\end{figure}

\begin{figure}
	\epsscale{1.2}
	\plotone{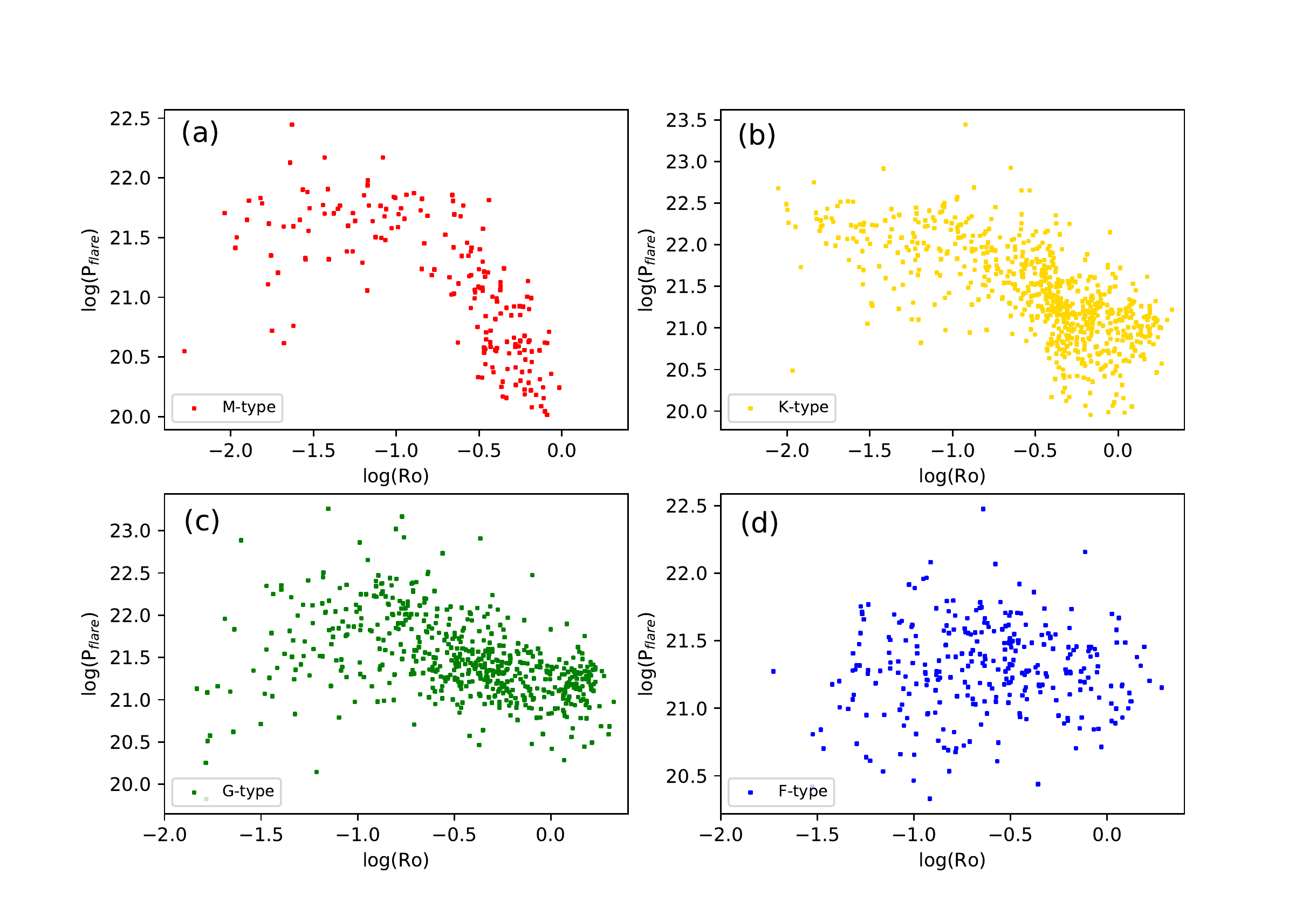}
	\caption{Total power of flares ($P_{\rm flare}$) versus logarithm of Rossby number for different spectral classifications: M-type stars (top left), K-type stars (top right), G-type stars (bottom left) and F-type stars (bottom right).}
	\label{Fig7}
\end{figure}

As we see from panel (b) of Fig \ref{Fig6} and \ref{Fig8}, both the frequency and total power of flares increase with increasing magnetic feature coverage and contrast. Before, the relationship between the spot area and flare activity has been studied extensively for G-type stars \citep{Notsu2013}. By studying flare characteristics of solar type stars, it is possible to evaluate how often can the Sun produce supeflare and the probability of such event when a big active region exists on the sun. For example, \cite{Maehara17} concluded that solar type stars showing superflares tend to have shorter rotation periods and larger starspot areas. Also the flare energy of solar type stars increases by increasing spot group area \citep{Notsu2013} and supports the idea that the upper limit of the energy released by the flares is consistent with the magnetic energy stored around the starspots \citep{Notsu2019}.

Since light curve fluctuations are higher in the case of one big starsspot in comparison with multiple starspots at different position of the stellar disk, the scatter of plots along horizontal axis in panel (b) of Figure \ref{Fig6} and Figure \ref{Fig8} refers to dispersion of $R_{eff}$, and could be correspond to various number of magnetic structures occupying different longitudes.

\subsubsection{Averaged Flux Magnitude of Flares} \label{subsubsec:Mf}
In Figure \ref{Fig9} averaged flux magnitude of flares ($M_{\rm flare}$) has been plotted versus different parameters. Panel (a) shows $M_{\rm flare}$ as a function of Rossby number that increases slowly by decreasing Rossby number especially in M and K-type stars. However, F-type stars have different behavior and their flares magnitude does not change dramatically when Ro decreases. This non-dependency to Rossby number may be a consequence of their possible different dynamo mechanism in which magnetic field production and the resulting reconnections is independent from rotation rate, see also panel (d) of Figure \ref{Fig2}, \ref{Fig3}. 
 
Averaged flux magnitude of flares does not show dependency to the magnetic feature stability (panel (b) of Figure \ref{Fig9}) in F-G type stars, however K-M stars show more dependency to the stability and their $M_{\rm flare}$ increases slightly by increasing $i_{AC}$. One possible explanation is that when magnetic features are more stable, magnetic energy may accumulate more over time, and the magnitude of the resulting flare from reconnection enhances. However, in hot stars with higher background intensity only huge flares can be detected and their magnitude dependency to different parameters can not be determined precisely.   

\begin{figure}
	\gridline{\fig{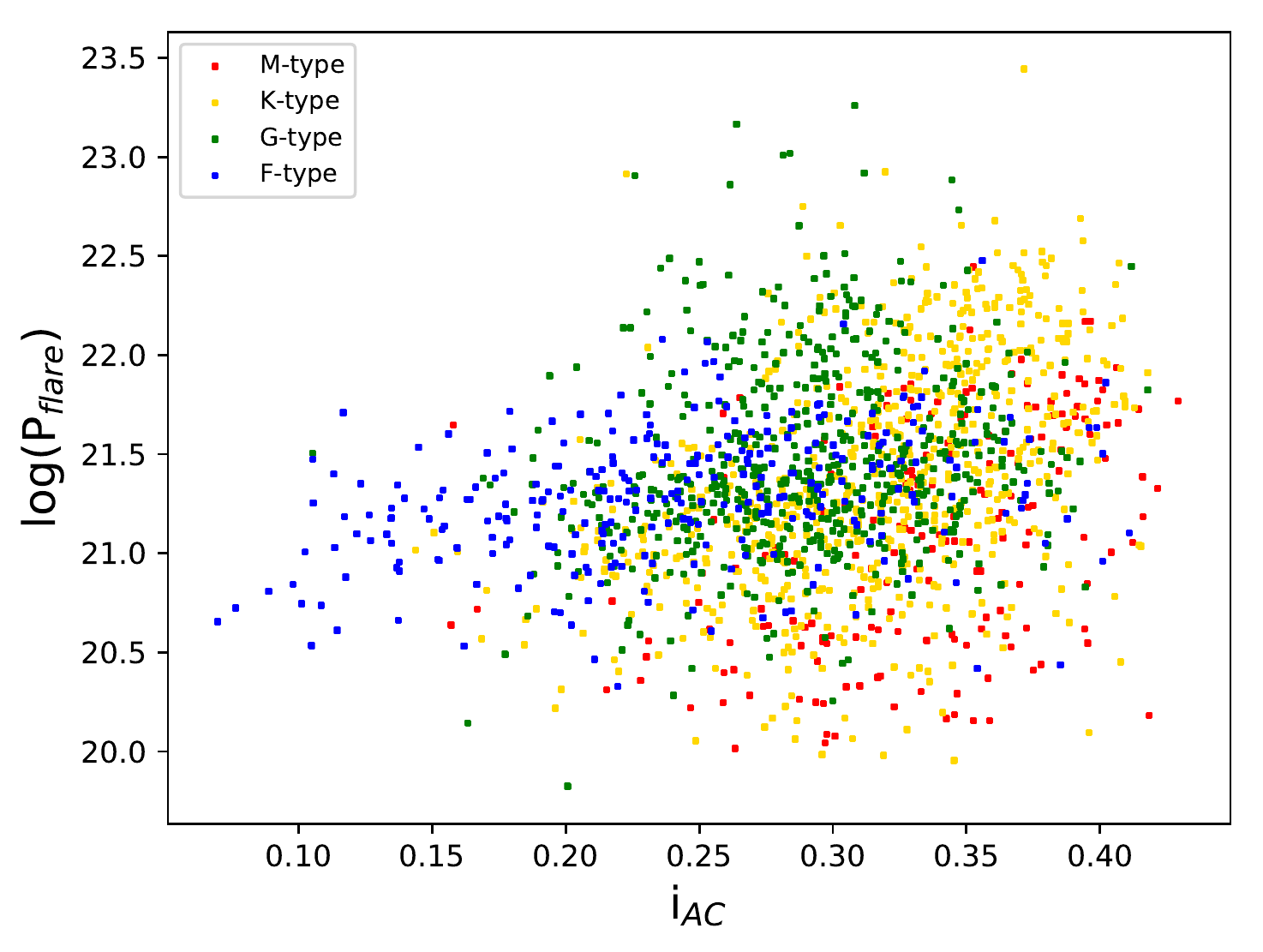}{0.51\textwidth}{(a)}
	\fig{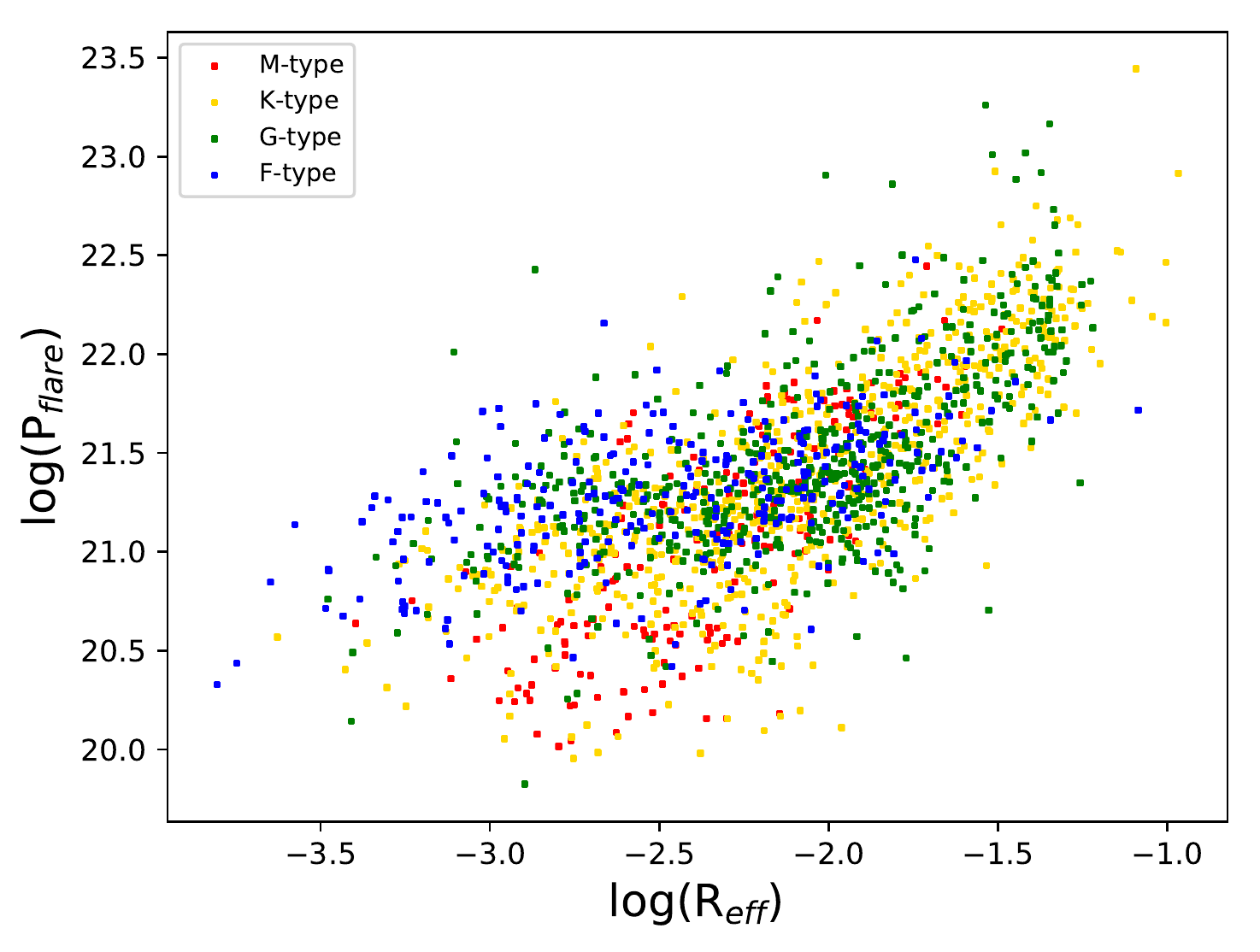}{0.51\textwidth}{(b)}}
	\caption{Variation of total power of flares ($P_{\rm flare}$) versus autocorrelation index ($i_{AC}$, left) and versus effective range of light curve fluctuations ($R_{eff}$, right) for different spectral types that are indicated with colors .}
	\label{Fig8}
\end{figure}

Generally, flare parameters seems to be more affected by the magnetic feature coverage and contrast rather than the stability. Figure \ref{Fig9}, panel (c) shows averaged flux magnitude of flares as a function of effective range of light curve fluctuations. $M_{\rm flare}$ increases by increasing $R_{eff}$ after $\log (R_{eff})=-2.0$ that is a critical threshold, see panel (b) of Figure \ref{Fig6} and Figure \ref{Fig8} that are $R_{\rm flare}$ and $P_{\rm flare}$ versus $R_{eff}$, respectively. By increasing $R_{eff}$ beyond this value, the slope of scatter plots suddenly enhances and flare frequency, power and magnitude improve dramatically. In other words, when relative coverage and contrast of magnetic structures become grater than a certain value, the probability of magnetic reconnections improves on one hand and the amount of magnetic energy stored within them increases on the other hand. This leads to higher flare frequency, power and magnitude.   

Figure \ref{Fig9} panel (d), shows $M_{\rm flare}$ versus time occupation ratio of flares ($R_{\rm flare}$) for different stars, averaged flux magnitude of flares increases in M-type stars by increasing $R_{\rm flare}$ and in G-K type stars increases slightly beyond $\log (R_{f})=-2.2$. Although variations of $M_{\rm flare}$ with $R_{\rm flare}$ is small, but we can conclude that the stars with higher flare frequency can also produce higher relative flux magnitude of flares. F-type stars do not show variations with $R_{\rm flare}$ and their flare rate does not have any effect on the flare magnitude. 

Overall behavior of the relative flux magnitude of flares indicates that in cool M and K-type stars $M_{\rm flare}$ is more affected by different parameters such as Rossby number, flare rate, magnetic feature stability, coverage and contrast than the hotter G-M type stars. This may arise from the different proto-conditions of the magnetic field production and its consequent reconnections during stellar dynamo. Especially the hottest group of the stars in this study (F-type) do no show any dependency to the mentioned parameter and their flare magnitude are rather constant. Nevertheless, the higher level of the background luminosity of the hot stars and the difficulty of detecting low magnitude flares should be taken into account. In cooler stars, low energy flares and the effect of different parameters are inferred easier.          

\subsection{Correlation Analysis} \label{subsubsec:corr}
In addition to median values of the parameters which were defined in section \ref{sec:param} and their results that were described in sections \ref{subsec:Magnetic} and \ref{subsec:flare}, correlation analysis is performed to study the possible relations between the magnetic and flare parameters of different quarters. This kind of correlation allows us to find out, whether the magnetic proxies and flare indexes are consistent with each-other. For each star, we calculate the correlation of pairs of parameter between different quarters, then the average values are computed (see table \ref{tab:corr}). Note that all of the correlation analysis are carried out between the quarters 2-16, so the results indicate that for each individual star whether the parameter values of different quarters correlate with each other or not. 

Figure \ref{Fig10} panel (a) shows the correlation between $i_{AC}$ and $R_{eff}$ for different stars as a function of Rossby number. The stability of magnetic features is positively correlated (mean value of 0.52) with their relative size and contrast \citep{Gnevyshev38, Petrovay97}. Bigger magnetic structures last more on the stellar surface following the same procedure that happen on the sunspots; the bigger the spots, the higher value of their lifetime \citep{Gnevyshev38, Petrovay97}. The value of correlation decreases by increasing temperature and maximum value belongs to M-type stars, see table \ref{tab:corr}. This may be the result of lower radius and background luminosity of the later type stars in which the effect of magnetic structures on the light curve variations could be more outstanding, so the correlation values are higher. 
 
\begin{figure}
	\epsscale{1.3}
	\plotone{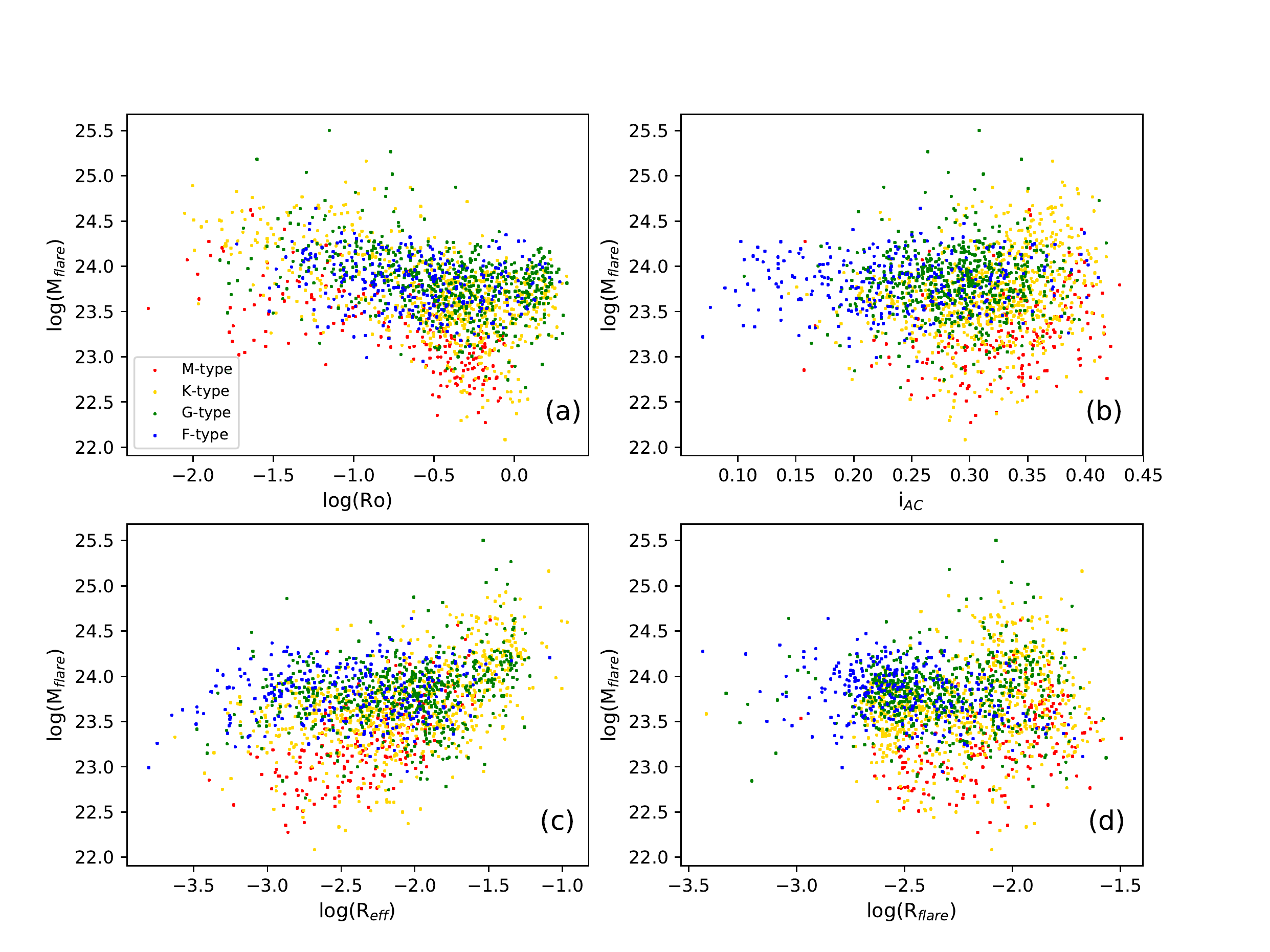}
	\caption{Variation of relative flux magnitude of flares ($M_{\rm flare}$) as a function of Rossby number (Ro, top left), autocorrelation index ($i_{AC}$, top right), effective range of light curve fluctuations ($R_{eff}$, bottom left) and flux magnitude of flares ($M_{\rm flare}$, bottom right) for different spectral types that are indicated with colors.}
	\label{Fig9}
\end{figure}

\begin{figure}
	\epsscale{1.3}
	\plotone{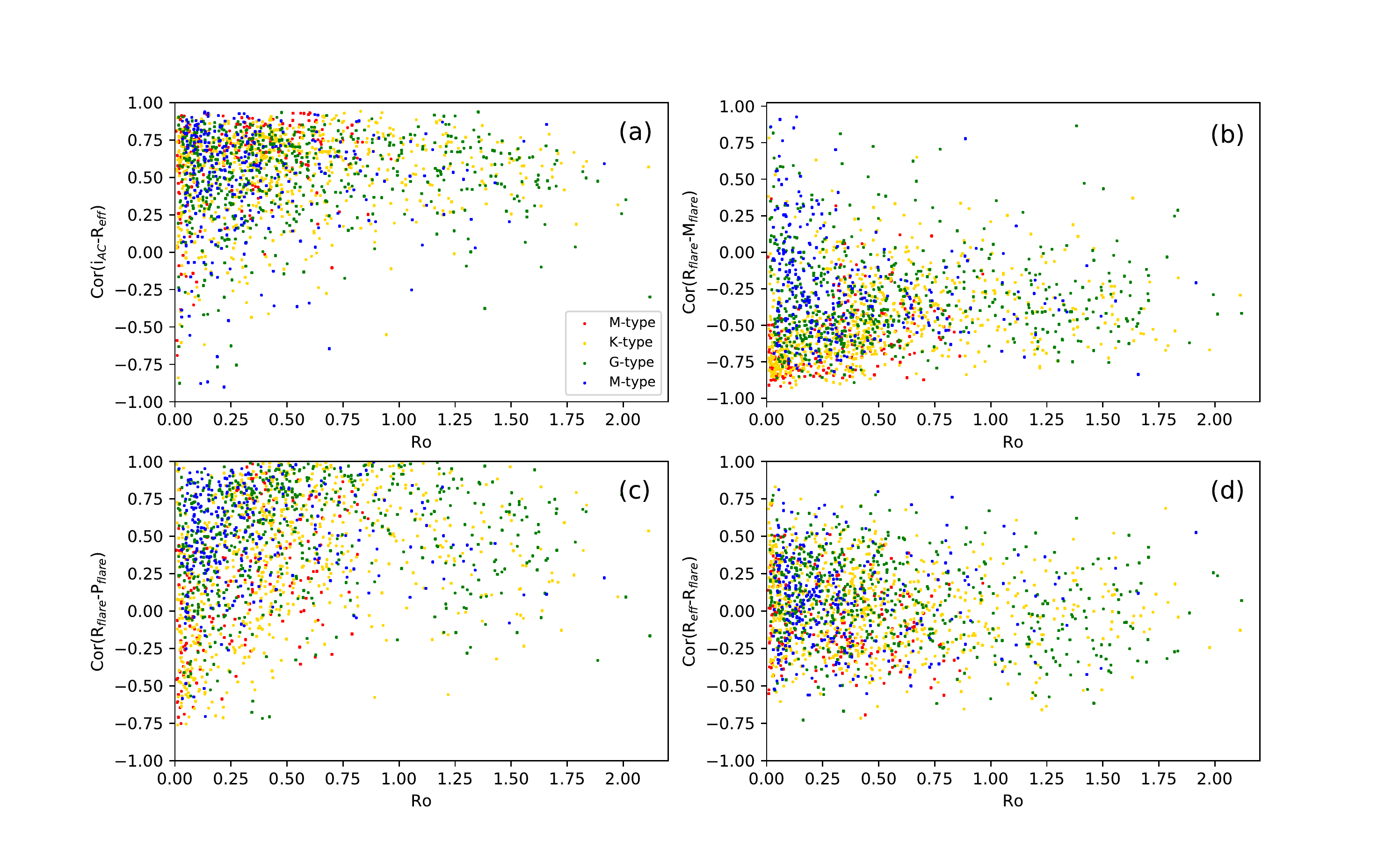}
	\caption{Average correlation of $i_{\rm AC}$ and $R_{\rm eff}$ (top left), $R_{\rm flare}$ and $M_{\rm flare}$ (top right), $R_{\rm flare}$ and $P_{\rm flare}$ (bottom left) and correlation of $R_{\rm eff}$ and $R_{\rm flare}$ (bottom right) between quarter 2-16 of each stars.}
	\label{Fig10}
\end{figure}

In Figure \ref{Fig10} panel (b), we see the correlation of time occupation ratio $R_{\rm flare}$ and averaged flux magnitude of flares ($M_{\rm flare}$) as a function of Rossby number. For most of the stars $R_{\rm flare}$ and $M_{\rm flare}$ are anti-correlated,  more occupied time of flares corresponds to the less averaged flux magnitude. This could mean that when several flares occur in the star, averaged flux magnitude decreases. In other words, magnetic energy releases during several flare events and prevents the accumulation of a huge magnetic energy which is required for the build up of major flares. 
By comparing with the result of Figure \ref{Fig9} panel (d) for K and M type stars, although the largest flares happen on the stars with highest flare rate, however the biggest flares do not necessarily coincide with the most flaring time of that star. The negative correlation of $R_{\rm flare}$ and $M_{\rm flare}$ is less prominent in the hot stars (see the table \ref{tab:corr}) possibly because of their higher background luminosity and the difficulty of measuring flare characteristics.  

Mean correlation of $R_{\rm flare}$ and $P_{\rm flare}$ is positive (mean value= 0.395) as expected, see Figure \ref{Fig10} panel (c). It means that a higher rate of flare activity corresponds to the higher total power, however some stars show negative values. These negatively correlated stars could be from two main groups, the first group produce a lot of (high time occupation ratio) small flares with low energy release. The second group produce few (low $R_{\rm flare}$) high energy flares. Generally, the correlation between $R_{\rm flare}$ and $P_{\rm flare}$ is grater in earlier type stars because the power is proportional with the temperature (see section \ref{sec:param}) and lower variation of flare frequency cause to the higher variation of produced power.
 
Other correlation analysis (correlation of $i_{AC}$-$R_{\rm flare}$; $i_{AC}$-$P_{\rm flare}$; $i_{AC}$-$M_{\rm flare}$; $R_{eff}$- $R_{\rm flare}$; $R_{eff}$- $P_{\rm flare}$; $R_{eff}$-$M_{\rm flare}$) do not show any positive or negative correlations (see table \ref{tab:corr}). For instance, one of them has been shown in Figure\ref{Fig10} panel (d), that is the correlation of $R_{eff}$ and $R_{\rm flare}$ which have mean value close to zero. It shows that the magnetic features which dominate the rotational modulation and the source region of the flares may not coincide at the same time. This is consistence with the result of \cite{Hawley2014} in which none of the energy and number of flares are correlated with the starspot phase. 

\floattable 
\begin{deluxetable}{lccccccccccccccc}
	\tablecaption{Average correlation values for pairs of parameters between quarter 2-16 of each spectral type.
		\label{tab:corr}}
	\tablehead{
		&&  \colhead{$i_{\rm AC}$-$R_{\rm eff}$}  & \colhead{$i_{\rm AC}$-$R_{\rm flare}$} & 
		\colhead{$i_{\rm AC}$-$P_{\rm flare}$}  & \colhead{$i_{\rm AC}$-$M_{\rm flare}$} & \colhead{$R_{\rm eff}$-$R_{\rm flare}$}
		&  \colhead{$R_{\rm eff}$-$P_{\rm flare}$}  & \colhead{$R_{\rm eff}$-$M_{\rm flare}$} &
		\colhead{$R_{\rm flare}$-$P_{\rm flare}$} & 	\colhead{$R_{\rm flare}$-$M_{\rm flare}$} \\                           
	}
	\startdata
	$T_{eff}<$3700 (M-type)  && 0.6030   & -0.0458 & -0.0605 & -0.0159 & -0.0568 & 0.0138 & 0.0407 & 0.2020 & -0.5773  &  \\
	3700$<T_{eff}<$5200 (K-type) && 0.5297 & -0.0150 & -0.0127& -0.0030 & 0.0196 & 0.0748 & 0.0376 & 0.3322 & -0.4665 &  \\
	5200$<T_{eff}<$6000 (G-type)  && 0.5055 & 0.0310 & 0.0244 & -0.00140 & 0.0750 & 0.0749& 0.0179&  0.4836 & -0.3215&  \\
	6000$<T_{eff}<$6700 (F-type) && 0.4754& 0.0043  & 0.0121 & 0.0050 & 0.0553 & 0.0917 & 0.0429& 0.4892&  -0.2425&  \\
	\enddata
\end{deluxetable}

\section{Conclusion} \label{sec:conclusion}
We analyzed the magnetic and flare activity of 1740 flare stars from Kepler long cadence data set. An automated algorithm was developed to extract the flare component from the light curve at a $3\sigma$ level as threshold value of the flare spikes. 
We studied different magnetic feature and flare  characteristics of the sample stars using two magnetic proxies and three flare indexes.
 
We defined two magnetic proxies, the first one is auto-correlation index ($i_{\rm AC}$) which represents the magnetic feature stability (see Figure \ref{Fig2}). The second one is effective range of the light curve fluctuation $R_{\rm eff}$, that reflects relative magnetic feature coverage and contrast and estimates the size of the largest magnetic features on the stellar surface. Both of $i_{\rm AC}$ and $R_{\rm eff}$ increase by decreasing Rossby number in G,K and M-type stars due to the more effective magnetic field production at higher rotation rates, until reaching to saturation level at $Ro\sim0.3$. In F-type stars, $i_{\rm AC}$ and $R_{\rm eff}$ do not show clear increase and saturation behavior and their lower average value of $i_{\rm AC}$ is due to the higher temperature and vigor of convection which results in greater turbulent diffusivity.   

Stellar magnetic features with bigger relative size and higher contrast are more stable than the smaller and lower contrast features (figure \ref{Fig4} and panel (a) of Figure \ref{Fig10}).
The relation between size and stability of magnetic features has been confirmed by different methods \citep{Namekata19} and also for solar magnetic structures like sunspots.

In general, dependency of the flare indexes such as time occupation ratio of flares ($R_{\rm flare}$), total power ($P_{\rm flare}$) and flux magnitude ($M_{\rm flare}$) on the magnetic feature coverage and contrast is more pronounced than their stability, see figure \ref{Fig6}, \ref{Fig8}. $R_{\rm flare}$ and $P_{\rm flare}$ increase by increasing $R_{\rm eff}$ for all spectral types of the sample because magnetic features with higher relative coverage and contrast have more potential to accumulate magnetic energy and suddenly release it in form of stellar flares after geometric rearrangement. 
 
The same as magnetic proxies, frequency and total power of flares also show an increase and saturation behavior. $R_{\rm flare}$ and $P_{\rm flare}$ increase by decreasing Rossby number due to excess of produced magnetic field from dynamo procedure until reaching to saturation level (around $Ro\sim0.18$ ) in M and K type stars (see Figure \ref{Fig3} and \ref{Fig5}). 
Low value of $R_{\rm flare}$ and $P_{\rm flare}$ at low Rossby numbers in some of the G and F-type stars can be due to their different mechanism of dynamo procedure, in which magnetic activity depends mainly to the temperature rather than the rotation rate \citep{Barnes2003,Yang2019}.
   
Average flux magnitude of flares increases by decreasing Ro and increasing $R_{\rm eff}$ (see fig \ref{Fig9}) in M, G and K-type stars.  $M_{\rm flare}$ does not change much with different parameters in F-type stars. However, it is worthwhile to mention that the difficulty of flare detection in hot stars with higher background luminosity may play role on their flare characteristics studies.    
 
The correlation analysis among quarters of each stars shows that when the frequency of flare events increases in the stars, averaged flux magnitude $M_{\rm flare}$ decreases (table\ref{tab:corr} and panel (b) of Figure \ref{Fig10}). When several flares occur in the stars, magnetic energy releases more often, so the huge magnetic energy accumulation needed for making big flares could not be built up.
Furthermore, the magnetic features that dominate the rotational modulation and the source region of the flares may not coincide with each-other (panel (d) of Figure \ref{Fig10}). Magnetic feature stability is not correlated with any of the flare parameters such as $R_{\rm flare}$, $P_{\rm flare}$ and $M_{\rm flare}$.

 
The magnetic analysis of the stars and their flare activity leads researchers to understand dynamo procedure at work beneath the stellar atmosphere and these results could help in advancing the theoretical magnetic field studies. In the future, by use of the observations from more advanced missions (such as TESS) a new era will be opened in stellar activity studies and the physics of magnetic structures .

\acknowledgments
This paper includes data collected by the Kepler mission. Funding for the \emph{Kepler} mission is provided by the NASA Science Mission directorate. The \emph{Kepler} data presented in this paper (\emph{Kepler} Data Release 25) were obtained from the Mikulski Archive for Space Telescopes (MAST). We thank referee for the critical comments and suggestions which helped us to improve our paper. 
H. He acknowledges the support of the Astronomical Big Data Joint Research Center, co-founded by the National Astronomical observatories, Chinese Academy of Sciences and the Alibaba Cloud.

\end{document}